\documentclass[]{interact}
\usepackage[ruled]{algorithm2e} 
\usepackage{mydefs}
\usepackage{url}
\usepackage{latexsym}
\usepackage{graphicx,color}
\usepackage[usenames,dvipsnames,svgnames,table]{xcolor}
\usepackage{bm}
\usepackage{bbm}
\usepackage{float}
\usepackage{soul}
\usepackage{enumitem}
\usepackage{subfigure}
\usepackage{lipsum}
\usepackage{booktabs}
\usepackage{parskip}
\usepackage{multirow}
\usepackage{tikz}
\usepackage{geometry}
\usepackage{pdflscape}
\usepackage{epsfig}
\usepackage{amsfonts}
\usepackage{latexsym}
\usepackage{epstopdf}
\usepackage{mathtools}
\usepackage{amsmath,amssymb,amsthm}
\usepackage{wrapfig}
\usepackage{comment}
\usepackage{colortbl}
\usepackage{multirow}
\usepackage{graphicx}
\usepackage{epstopdf}% To incorporate .eps illustrations using PDFLaTeX, etc.
\usepackage[caption=false]{subfig}% Support for small, `sub' figures and tables
\usepackage[normalem]{ulem}
\usepackage{soul}

\usepackage[numbers,sort&compress]{natbib}% Citation support using natbib.sty
\bibpunct[, ]{[}{]}{,}{a}{,}{,}% Citation support using natbib.sty
\renewcommand\bibfont{\fontsize{10}{12}\selectfont}% Bibliography support using natbib.sty

\theoremstyle{plain}% Theorem-like structures provided by amsthm.sty

\theoremstyle{definition}

\theoremstyle{remark}

\graphicspath{{figures/}}

\begin{document}

\title{Semiparametric Latent ANOVA Model for Event-Related Potentials}

\author{
\name{Cheng-Han Yu\textsuperscript{a}\thanks{CONTACT Marina Vannucci. Email: marina@rice.edu}, Meng Li\textsuperscript{b}, and Marina Vannucci\textsuperscript{b}}
\affil{\textsuperscript{a}Department of Mathematical and Statistical Sciences, Marquette University, Milwaukee WI, USA; \textsuperscript{b}Department of Statistics,  Rice University, Houston, TX, USA}
}

\maketitle

\begin{abstract}
Event-related potentials (ERPs) extracted from electroencephalography (EEG) data in response to stimuli are widely used in psychological and neuroscience experiments. A major goal is to link ERP characteristic components to subject-level covariates. Existing methods typically follow two-step approaches, first identifying ERP components using peak detection methods and then relating them to the covariates. This approach, however, can lead to loss of efficiency due to inaccurate estimates in the initial step, especially considering the low signal-to-noise ratio of EEG data. To address this challenge, we propose a semiparametric latent ANOVA model (SLAM) that unifies inference on ERP components and their association to covariates. SLAM models ERP waveforms via a structured Gaussian process prior that encodes ERP latency in its derivative and links the subject-level latencies to covariates using a latent ANOVA. This unified Bayesian framework provides estimation at both population- and subject- levels, improving the efficiency of the inference by leveraging information across subjects. We automate posterior inference and hyperparameter tuning using a Monte Carlo expectation-maximization algorithm. We demonstrate the advantages of SLAM over competing methods via simulations. Our method allows us to examine how factors or covariates affect the magnitude and/or latency of ERP components, which in turn reflect cognitive, psychological or neural processes. We exemplify this via an application to data from an ERP experiment on speech recognition, where we assess the effect of age on two components of interest.  Our results verify the scientific findings that older
people take a longer reaction time to respond to external stimuli because of the delay in perception and brain processes.
\end{abstract}


\section{Introduction}
Event-related potentials (ERPs) refer to the electrical potentials measured by extracting voltages from electroencephalography (EEG) data in response to internal or external stimuli, typically during visual, auditory, or gustatory experiments. Raw EEG signals summarize all neural activity in the brain, making it difficult to distinguish specific neural processes related to perception, cognition, and emotion. In contrast, ERPs can identify multiple neurocognitive and affective processes contributing to behavior \citep{Gasser1996}. With other advantages such as noninvasiveness, continuous high temporal resolution, and relatively low cost compared to other neuroimaging modalities such as functional magnetic resonance imaging, ERPs have become a widely used tool in psychological and neuroscience research \citep{Vogel2004, Luck2022}.

Clinical studies usually focus on specific characteristic components of the ERP waveforms, defined as voltage deflections produced when specific neural processes occur in specific brain regions \citep{Luck2012}. A main task of any ERP analysis is to estimate the amplitude, i.e. magnitude (in microvolts, $\mu$V) and the latency, i.e. position in time (in milliseconds, ms) of specific ERP components. In current ERP studies, most often researchers identify such components by visually inspecting the \textit{grand} ERP waveform averaged across trials and subjects,  often relying on previous scientific research findings. When the statistical analysis is done at the subject level, two-step approaches are typically used, first extracting the ERP components of interest and then performing an analysis of variance (ANOVA) to relate the extracted components to the subject-level covariates. In such analyses, in order to obtain a smooth ERP curve, the whole EEG experiment needs to be averaged across many trials and/or participants. In addition, filtering is necessary to remove the signal trend or drifts, together with noise that comes from biological artifacts or the electrical activity recording environment. Once a smoothed ERP curve has been obtained, a time window needs to be specified as the range of the component of interest. This is a critical choice in many of the analyses. For example, one of the common ways to quantify the amplitude of an ERP component of interest is as the mean amplitude integrated over the chosen window. This may lead to false discoveries when comparing the amplitudes of ERP components across control and experimental groups.

In statistics, there is limited contribution to methods for the identification and estimation of ERP components. \cite{Jeste2015} and \cite{hasenstab2015identifying} use LOESS-based methods that first smooth the noisy waveforms and then apply a peak detection algorithm on the smoothed curve to identify optima within a specified time window. \cite{hasenstab2015identifying}, in particular, propose MAP-ERP, a meta-preprocessing step based on a moving average of the ERP functions over trials in a sliding window, to preserve the longitudinal information in ERP data, and then employ peak detection. With peak detection algorithms, latency is calculated as the location of the stationary point identified within a pre-set window and the amplitude is calculated as the value of the smoothed curve at the location of the stationary point or by integrating the area under the curve. Other approaches for detecting ERP components, in single subjects, use continuous wavelet transformation techniques, see for example \cite{Kalli}.  

Recently, various extensions of Gaussian Processes (GPs) have been applied to neuroscience problems. \cite{Kang2018} propose a soft-thresholded GP prior for feature selection in scalar-on-image regression utilized to represent sparse, continuous, and piecewise-smooth functions. \cite{Yu2022} employ Gaussian processes with derivative constraints, defining a peak or dip location as the stationary point such that the first derivative of the underlying smooth ERP waveform at the point is zero. While this method takes the data averaged across trials for analysis, as traditional studies do, it does not need separate smoothing steps of the ERP waveforms. \cite{Ma2023} develop the split-and-merge GP prior for selecting time windows in which the P300 ERP components in response to the target and non-target stimuli may be different. While the model considers multiple channel EEG signals and spatial correlation, it focuses on the amplitude difference and uncertainty about ERP functions.

In this paper we propose a novel semiparametric latent ANOVA model (SLAM) as a data-driven model-based approach that provides a unified framework for the estimation of ERP characteristics components and their association to subject-level covariates. Following \cite{Yu2022}, we employ a Bayesian model with derivative-constrained Gaussian process priors, to produce smooth estimates of the ERP waveform together with amplitude and latency of ERP components, and incorporate into the model a flexible ANOVA setting, to examine how multiple factors or covariates, such as gender and age, affect the ERP components' latency and amplitude. Unlike the model of \cite{Yu2022}, which considers one single ERP waveform at a time, either as an averaged single-subject ERP or a grand mean across subjects, our hierarchical framework allows estimates of latency and amplitude at the group level as well as at the individual subject level. To produce such inference, SLAM starts with an initial partition of the experimental time interval into sub-intervals, one for any potential ERP component location, and then employs derivative-constrained GPs to identify the latency of the components. The posterior distribution of latency parameters not only provides a more precise identification of the ERP component location, but allows the calculation of credible intervals that can be used as a refined search window where to compute amplitude estimates or, indirectly, as a reference in follow-up studies, to corroborate previous scientific findings or experts' knowledge. We perform posterior inference and hyperparameter tuning via a Monte Carlo expectation-maximization algorithm. Through simulations, we demonstrate how our proposed method produces smaller root mean square error for both latency and amplitude estimates, compared to the LOESS-based method. Unlike these methods, our approach provides a direct estimate of the latency parameter, without the need to apply curve smoothing. Moreover, SLAM provides automatic group-level estimates, while an additional ANOVA is needed when the LOESS-based methods are used.

In neuroscience and psychological sciences, ERP components are typically used to examine perception or cognitive differences between control and experimental groups or under different levels of an influential factor. In fact, the components are usually named by whether it is a positive wave (peak) or a negative wave (dip), and when it occurs. For example, the peaks occurring around 100 msec, N100, or around 200 msec, P200, after the start of the stimulus, have been implicated in speech recognition \citep{Noe2019}, the N170 component in object recognition \citep{Tanaka2001} and the N400 in language processing \citep{Kutas1980}. Here we consider ERP data collected during an experiment designed to determine whether early stages of speech perception are independent of top-down influences.  Our hierarchical framework allows us to estimate latency at both subject- and group-level. Our results show that the older group people tend to have longer latency of both the N100 and P200 ERP components. We also find that the older group people tend to have larger N100 amplitude. These results verify the scientific findings that older people take a longer reaction time to respond to external stimuli because of the delay in perception and brain processes \citep{Noe2019, Yi2014, Tremblay2003}.

The rest of the paper is organized as follows: Section \ref{sec:slam} describes our proposed Bayesian hierarchical model and its associated algorithm for learning ERP waveforms and components. Section \ref{sec:sim} provides a simulation study, showing the estimation performance of the proposed method against LOESS-based approaches, and Section \ref{sec:erp} illustrates the application of our methodology to an ERP data set. Section \ref{sec:conclude} presents our conclusions. 

\section{Methods}
\label{sec:slam}
In this section, we describe the proposed general SLAM framework, for the analysis of multi-subject-multi-group ERP data. To ease exposition, we consider a one-way ANOVA setting, and will use \textit{age} (``old'' and ``young'') as demonstrative example to illustrate our approach. We note, however, that the model construction can be easily generalized to incorporate any given number of factors and continuous covariates.

\subsection{Derivative-Constrained Gaussian Process}

Let $y_{igs}$ be the observation at time $x_i \in \mathcal{X}$, for $i = 1, \ldots, n$, with $g = 1, 2, \dots, G$ indicating the levels of a factor, and with $s = 1, \dots, S_{g}$ indexing the subject within the $g$th group. We assume the following model: 

\begin{equation}\label{mdl:lik}
y_{igs} = f_{gs}(x_i) + \epsilon_{igs}, ~~ \epsilon_{igs} \iid N(0, \sigma^2).
\end{equation}
In ERP applications,  $f_{gs}(\cdot)$ represents the underlying true ERP waveform at level $g$ of the factor, e.g., the ERP waveform for the older adult group. This waveform is usually believed to be a smooth curve.

We adopt the Bayesian modeling approach of \cite{Yu2022} and impose a derivative-constrained Gaussian Process prior on the function $f_{gs}(\cdot)$ \citep{rasmussen2006,Ghosal17}. We assume that the function has $M$ stationary points, $\bm{t}_{gs} = \{t_{gs}^m\}_{m = 1}^M,$ each corresponding to the latency of an ERP component. The parameters $\bm{t}_{gs}$ and their association with subject-level covariates are of interest here.  We encode the stationary points $\bm{t}_{gs}$ in the modeling of the ERP waveforms via a derivative Gaussian process (DGP) as
\begin{equation}
    f_{gs} \mid \bm{t}_{gs} \sim \text{DGP}(0, k(\cdot, \cdot; \btheta), \bm{t}_{gs}), 
\end{equation}
where $\text{DGP}(0, k(\cdot, \cdot; \btheta), \bm{t}_{gs})$ is a GP with mean 0 and covariance kernel $k(\cdot, \cdot; \btheta)$ conditioned upon $f_{gs}'(t_{gs}^m) = 0$ for $m = 1, \ldots, M$, which is also a GP with $\bm{t}_{gs}$-dependent mean and covariance functions. Here $\btheta$ indicates the hyperparameters in the kernel.  Gaussian process priors are widely used for modeling unknown functions, and the covariance kernel is instrumental in determining sample path properties. \cite{Yu2022} and \cite{Li2023} show that a general $\text{DGP}$ forms a $n$-dimensional Gaussian distribution $N\left(\bmu_{gs}, \bSig_{gs}(\btheta, \bm{t}_{gs})\right)$ with mean vector 
$$\bmu_{gs} = \mu_{gs}(\bm{x}) - k_{01}(\bm{x}, \t_{gs})k_{11}^{-1}(\t_{gs}, \t_{gs})\mu'_{gs}(\t_{gs}),$$ and covariance matrix 
$$\bSig_{gs} = k_{00}(\bm{x}, \bm{x}) - k_{01}(\bm{x}, \t_{gs})k_{11}^{-1}(\t_{gs}, \t_{gs})k_{10}(\t_{gs}, \bm{x}),$$ where $k_{00}$ is the $n \times n$ covariance matrix of $\bm{x}$, $k_{10} = k_{01}'$ is the $M \times n$ covariance matrix of $\bm{x}$ and $\t_{gs}$, and $k_{11}$ is the $M \times M$ covariance matrix of $\t_{gs}$.
Following these authors, we use the squared exponential kernel, that is 
\begin{equation}
k(x, x'; \btheta = (\tau, h)) = \tau^2\exp\left(-\frac{1}{2h^2} \|x - x'\|^2 \right),
\end{equation}
as the waveform in ERP studies is typically believed to be smooth.

In the absence of prior information on $M$, the number of stationary points, the method developed in \cite{Yu2022} and \cite{Li2023} can be used to estimate $M$ and provide interval estimates for each latency.  Specifically, $M$ is estimated by the number of disjoint segments in the highest posterior density region of the posterior distribution of stationary points. In many ERP applications, however, the number $M$ can be easily pre-specified at a small fixed value, say one to three, based on prior knowledge of the process under study, and a time interval $[a^{m}, b^{m}]$ can be specified as the search window of interest for the $m$-th ERP component before building the model. 
When only vague information is available, the search windows could form a partition of the entire EEG experiment epoch, that is, $[a^{m}, b^{m}] \cap [a^{m'}, b^{m'}] = \emptyset$ for $m \ne m'$ and $\cup_{m=1}^M [a^{m}, b^{m}] = \mathcal{X}$. Note that $a^{m}$ and $b^{m}$ can be specified in a factor-dependent manner.

\subsection{Latent ANOVA Model}
Unlike \cite{Yu2022}, which focused on employing DGP to infer the unknown $M$ components on a single ERP waveform, our goal is to model ERP components across multiple subjects and examine how multiple factors or covariates, such as gender and age, affect the ERP components’ latency and amplitude. For this, we incorporate a flexible ANOVA construction into our model, that allows estimation of the ERP components at both the population and subject-specific levels.

For each level $g$, the subject-level parameter $t_{gs}^m$ is governed by the general beta prior supported on $[a^{m}, b^{m}]$ 
\begin{align}\label{eqn:t}
t_{gs}^m \mid r_{g}^m, \eta_{g}^m &\iid \text{gbeta}\left(r_{g}^m\eta_{g}^m, \, (1 - r_{g}^m)\eta_{g}^m, \, a^{m}, \, b^{m}\right), \, s = 1, \dots, S_{g}. 
\end{align}

The general beta distribution supported on $[a, b]$ is distributed as $(b - a) X + a$ for the standard beta variable $X$ supported on [0, 1]. The first two parameters in the general beta distribution are parametrized such that $r_{g}^m \in (0, 1)$ is the location parameter and $\eta_{g}^m$ is the scale parameter. The prior mean of $t_{gs}^m$ is $\ex \left(t_{gs}^m\right) = (1 - r_{g}^m)a^{m} + r_{g}^mb^{m}$. Note that all subjects in the same level $g$ have the $m$-th stationary point from the common level general beta distribution centered at $r_{g}^m$ in its time window. Such hierarchy is a starting point of the ANOVA structure as the mean $r_{g}^m$ is then further decomposed into the overall mean and the main effect from the factor.

The group-level parameter $r_{g}^m$ indicates the relative location of the mean in the interval $[a^{m}, b^{m}]$, or the weight between the two end points of the search window. The larger $r_{g}^m$ is, the closer the prior mean is to $b^{m}$. The weight $r_{g}^m$ can be easily transformed to have the same scale as $t_{gs}^m$ and serves as the group-level latency that is common across subjects at the same factor levels. Given $r_{g}^m$, the value of $\eta_{g}^m > 0$ controls the prior variance of $t_{gs}^m$, which is $\var\left(t_{gs}^m\right) = \dfrac{(b^{m} - a^{m})^2r_{g}^m(1-r_{g}^m)}{1+\eta_{g}^m}$. One can specify $\eta_{g}^m$ using fixed values based on domain knowledge or preliminary estimates, or apply a common prior for within-group variability.

For inference on $r_{g}^m$, we introduce an ANOVA structure to model how the two factors affect the location of the ERP component. Since $r_{g}^m$ is bounded between zero and one, we use a link function $\phi\left( \cdot \right) \in (-\infty, \infty)$ followed by an ANOVA:

\begin{equation}\label{eqn:anova}
\phi\left(r_{g}^m\right) = \beta_0^m + \beta_{1}^mz_{1} + \cdots + \beta_{G-1}^mz_{G-1}.
\end{equation}

Here we write the one-way ANOVA as a linear model using dummy variables $\left\{z_{a}\right\}_{a=1}^{G-1}$ where $z_{a} \in \{0, 1\}$ is a binary variable denoting the level of factor. Coefficient $\beta_0^m$ denotes the grand mean at the baseline level of factor, and $\beta_{a}^m$ quantify the effect of level $a$ comparing to the baseline.

This latent structure is flexible and inclusive in the following sense. First, it can accommodate any linear model with categorical and numerical variables. For example, subject-level numerical covariates (eg, blood pressure) can be included in the latent regression. Second, the latent structure is able to accommodate interactions between factors. Moreover, any link function for parameter values restricted in $(0, 1)$ can be used in the modeling setup, such as logit, probit, and complementary log-log links.

\subsection{Complete Model}
We complete the model by placing standard normal priors on the beta coefficients in (\ref{eqn:anova}), a gamma prior on the positive variable $\eta_{g}^m$ and standard inverse gamma priors on $\sigma^2$. In summary, the full Bayesian SLAM is summarized as follows. For time point $i = 1, \dots, n$, subject $s = 1, \dots, S_{g}$, factor level $g = 1, \dots, G$, and stationary point index $m = 1, \dots, M$,
\begin{align*}
y_{igs} &= f_{gs}(x_i) + \epsilon_{igs}, ~~ \epsilon_{igs} \iid N(0, \sigma^2),\\
f_{gs}(\cdot) &\sim GP(0, k(\cdot, \cdot; \tau, h)),\\
f_{gs}'(t_{gs}^m) &= 0, \\
t_{gs}^m \mid r_{g}^m, \eta_{g}^m &\iid \text{gbeta}\left(r_{g}^m\eta_{g}^m, (1 - r_{g}^m)\eta_{g}^m, \, a^{m}, \, b^{m}\right), \\
\phi\left(r_{g}^m\right) &= \beta_0^m + \sum_{a=1}^{G-1}\beta_{a}^mz_{a}\\
\beta_0^m &\sim N(\mu_0^m, (\sigma_0^m)^2),\\
\beta_{a}^m &\sim N(\mu_1^m, (\sigma_1^m)^2),\\
\eta_{g}^m &\sim Ga(\alpha_{\eta}, \beta_{\eta}),\\
\pi(\sigma^2, \tau^2, h) &= \pi(\sigma^2) \pi(\tau^2) \pi(h) = IG(\alpha_{\sigma}, \beta_{\sigma})IG(\alpha_{\tau}, \beta_{\tau})Ga(\alpha_{h}, \beta_{h}).
\end{align*}

With no prior knowledge, we suggest setting hyperparameters $\mu_j^m, j = 0, 1$ at zero, and $(\sigma_j^m)^2, j = 0, 1$ at one, i.e., standard Gaussian distribution. In our experience, the parameter learning is insensitive to the variance size. Furthermore, to circumvent overfitting, we choose hyperparameters in the inverse gamma prior on $\sigma^2$ so that its mean is around the empirical mean of the data. As for the hyperparameters $\tau^2$ and $h$ in the covariance function, we follow \cite{Yu2022} and estimate then based on the marginal maximum likelihood (MML) by embedding an EM algorithm into the posterior sampling, as explained in the section below, therefore avoiding a separate tuning of these hyperparameters.

\subsection{Posterior Inference}
We adopt the Monte Carlo expectation-maximization (MCEM) algorithm for joint parameter tuning and posterior sampling of \cite{Yu2022} to our SLAM framework.  More specifically, in the E-step, we sample the subject-level latency parameters, $t_{gs}^m$, the factor main effect for group-level latencies, $\beta_0^m, \beta_{a}^m$, the general beta scale parameters, $\eta_{g}^m$, and the noise variance, $\sigma^2$,  and optimize the hyperparameters $\tau^2$ and $h$ in the M-step. 

Let $\y = (\y_{1}, \dots, \y_{G})$, with $\y_{g} = (\y_{g1}, \dots, \y_{gS_{g}})$ and $\y_{gs} = (y_{1gs}, \dots, y_{ngs})$. Similarly, let $\t = (\t_{1}, \dots, \t_{G})$, with $\t_{g} = (\t_{g}^1, \dots, \t_{g}^M)$, $\t_{gs} = (t_{gs}^1, \dots, t_{gs}^M)$, $\t_{g}^m = (t_{gs}^m, \dots, t_{gS_{g}}^m)$,  and $\r = (\r_{1}, \dots, \r_{G})$, with $\r_{g} = (r_{g}^1, \dots, r_{g}^M)$. Let us also define $\bbeta = (\bbeta_0, \bbeta_1)$, with $\bbeta_0 = (\beta_0^1, \dots, \beta_0^M)$,  $\bbeta_1 = (\bbeta_1^1, \dots, \bbeta_1^M)$, $\bbeta_1^m = (\beta_{1}^m, \dots, \beta_{G-1}^m)$, $\bet = (\bet_{1}, \dots, \bet_{G})$, with $\bet_{g} = (\eta_{g}^1, \dots, \eta_{g}^M)$, and $\btheta = (\tau, h)$.  At iteration $j$, the posterior density for the E-step is 
\begin{align*}
	q\left(\t, \bbeta, \bet, \sigma^2 \mid \y, \hat{\btheta}^{(j)} \right) &\propto p\left(\y \mid \t, \bbeta, \bet, \hat{\btheta}^{(j)} \right) \pi\left(\t, \bbeta, \bet, \sigma^2\right) \\
	& \propto p\left(\y \mid \t, \hat{\btheta}^{(j)} \right)\pi\left(\t \mid \r(\bbeta), \bet \right) \pi(\bbeta) \pi(\bet) \pi(\sigma^2),
\end{align*}
and, for each $g$ and $s$, the marginal likelihood with $f_{gs}$ being integrated out is 
$$p\left(\y_{gs} \mid \t_{gs}, \hat{\btheta}^{(j)} \right) = \int p\left(\y_{gs} \mid f_{gs}, \t_{gs}, \hat{\btheta}^{(j)} \right) p(f_{gs}) \, d \,f_{gs},$$ 
which is a $n$-dimensional Gaussian distribution
$$\y_{gs} \mid \t_{gs}, \hat{\btheta}^{(j)} \ind N\left(\bmu_{gs}, \bSig_{gs} +\sigma^2\I \right),$$ where $\bmu_{gs}$ and $\bSig_{gs}$ are as defined in Section 2.1. The full conditional distributions for all parameters can be found in the Supplementary Material. 

Given the posterior samples of $\t$ and $\sigma^2$ drawn from the E-step, the hyperparameters $\tau^2$ and $h$ are updated in the M-step by maximizing the log marginal conditional likelihood:
\begin{equation} \label{eqn:theta}
\hat{\btheta}^{(j+1)} = \arg \max_{\btheta} \sum_{l=1}^L \log  p\left(\y \mid \t_l^{(j)}, (\sigma^2_l)^{(j)}, \hat{\btheta}^{(j)} \right),
\end{equation}
where  $\t_l^{(j)}$ and $(\sigma^2_l)^{(j)}$ are the $l$-th posterior sample of $\t$ and $\sigma^2$ at the $j$-th iteration in the algorithm respectively. Therefore, $\hat{\btheta}^{(j+1)}$ is the maximum marginal likelihood (MML) estimates at $(j+1)$-th iteration in the MCEM algorithm. The optimization step can be solved numerically using standard numerical optimization approaches.  The detailed sampling scheme in the MC E-step is discussed in the Appendix, and the entire algorithm is summarized in Algorithm \ref{algo:MCEM} below. 

The MCEM reaches convergence when $\| \hat{\btheta}^{(j+1)} - \hat{\btheta}^{(j)} \| < \epsilon$ at the $(j+1)$-th iteration \citep{Bai2019}. The threshold $\epsilon$ is set to be $10^{-5}$ in the simulation and data analysis. With the sample size of 100, our R code takes about 70 seconds for $J=500$ to finish one M-step, and 30 to 35 seconds for drawing 1000 samples in the MC E-step or final simulation.

\begin{algorithm}
	\fontsize{12}{6}\selectfont
	\SetAlgoLined
	\textbf{Initialize:}\\
    1. Set model values of $\epsilon$, $D$, $G$, $L$, $M$, $a^{m}$, $b^{m}$, $\alpha_{\sigma}$, $\beta_{\sigma}$, $\alpha_{\eta}$, $\beta_{\eta}$, $\mu_0^m$, $\mu_1^m$, $\delta_0^m$, $\delta_1^m$,  $\alpha_{\tau}$, $\beta_{\tau}$,$\alpha_{h}$, $\beta_{h}$\\
	2. Set initial values for the algorithm: $\hat{\btheta}^{(1)}$, $(t_{gs}^m)^{(0)} \in[a^{m}, b^{m}]$, $(r_{g}^m)^{(0)} \in(0,1)$,  $(\eta_{g}^m)^{(0)} > 0$, $(\sigma^2)^{(0)}>0$.\\
	\vskip 2mm
	3. For iteration $j = 1, 2, \dots$ until convergence criterion $\|\hat{\btheta}^{(j)} - \hat{\btheta}^{(j+1)}\| < \epsilon$ is satisfied:\\
	\vskip 2mm
	
	\textbf{Monte Carlo E-step:}\\ 
	\vskip 1mm
		
		\For{$d=1, \dots, D$}{
		For each $g$ and $s$, draw $\t_{gs}^{(j)}$ in blocks using Metropolis-Hasting steps on its conditional density $p\left(\t_{gs} \mid \y_{gs}, \r_{g}, \sigma^2, \hat{\btheta}^{(j)} \right)$\\
		For each $m$, draw $(\beta_{0}^m)^{(j)}$ using Metropolis-Hasting steps on its conditional density $p\left(\beta_0^{m} \mid \t, \bet \right)$\\
		
		For each $a$ and $m$, draw $(\beta_{a}^m)^{(j)}$ using Metropolis-Hasting steps on its conditional density $p\left(\beta_{a}^{m} \mid \t, \bet \right)$\\
					
		Obtain $\phi\left((\beta_{0}^m)^{(j)}, (\beta_{a}^m)^{(j)}\right)$ and use the inverse link to transform $\phi$ to $(r_{g}^m)^{(j)}$
		
		For each $g$ and $m$, draw $(\eta_{g}^m)^{(j)}$ using Metropolis-Hasting steps on its conditional density $p\left(\eta_{g}^m \mid \t_{g}^m, r_{g}^m \right)$ \\
		Draw $(\sigma^2)^{(j)}$ from its inverse gamma conditional distribution $IG\left(\frac{n}{2}(\sum_{g=1}^G S_{g}) + \alpha_{\sigma}, \left(\frac{1}{2}\sum_{g=1}^G\sum_{s=1}^{S_{g}}\y_{gs}'A_{gs}^{-1}\y_{gs}\right)+\beta_{\sigma}\right)$\\
		$d \leftarrow d + 1$
		}

	\vskip 2mm
	Draw sample $\{(t_{gs, l}^m)^{(j)}\}_{l = 1}^L$ from $\{(t_{gs, d}^m)^{(j)}\}_{d = 1}^D$.\\
	Draw sample $\{(\sigma_l^2)^{(j)}\}_{l = 1}^L$ from $\{(\sigma_d^2)^{(j)}\}_{d = 1}^D$. 
	\vskip 2mm
	\textbf{M-step:}\\ Given samples of $\t$ and $\sigma^2$ at the $j$-th iteration,  $\{(t_{gs, l}^m)^{(j)}\}_{l = 1}^L$ and $\{(\sigma_l^2)^{(j)}\}_{l = 1}^L$, update $\btheta$ to $\hat{\btheta}^{(j+1)}$ by optimizing
	$\hat{\btheta}^{(j+1)}$ according to eq. (\ref{eqn:theta}).
	\caption{Monte Carlo EM Algorithm for SLAM}
	\vskip 2mm
	\KwResult{MML estimate of $\btheta$ and posterior samples of $\t$, $\bbeta$, $\bet$ and $\sigma^2$}
	\label{algo:MCEM}
\end{algorithm}

For multi-way ANOVA, for example with two factors, A with levels $g = 1, \dots, G$ and B with levels $h = 1, \dots, H$, the latent ANOVA structure can be directly extended into

$$\phi\left(r_{gh}^m\right) = \beta_0^m + \sum_{a=1}^{G-1}\beta_{1, a}^mz_{1, a}+ \sum_{b=1}^{H-1}\beta_{2, b}^mz_{2, b}$$
where $z_{1, .} \in \{0, 1\}$ is a binary variable for factor A, and $z_{2, .} \in \{0, 1\}$ for factor B. The posterior inference and learning algorithm are similar to the one-way ANOVA demonstration.

\section{Simulation Study}
\label{sec:sim}
In this section, we examine the performance of our proposed method through a simulation study, where we also compare results with the LOESS-based method of \cite{Jeste2015}. For latency inference, we also discuss the difference between SLAM and the DGP-MCEM method of \cite{Yu2022}.

\subsection{Data Generation} \label{sec:simdata}
For a fair comparison of the proposed method with other competing methods, in this section we consider a simulation setting where the data are not generated directly from the proposed model. We show that our model estimates the latency location better than the other methods. Later in Section \ref{sec:simdatamodel}, we examine how well the model learns the parameters in a scenario where the data are generated from our model. We consider the case of one factor with two levels or groups, each having ten subjects and ERP waveforms. Depending on how smooth or noisy the data set is, the ten ERP signals could be treated as either waveforms of ten trials of a single subject or waveforms averaged across trials for ten subjects. The two groups have significantly different ERP patterns, while the individuals within groups behave similarly with little latency and/or amplitude shifts.

\begin{figure}
\centering
\includegraphics[width=2.5in]{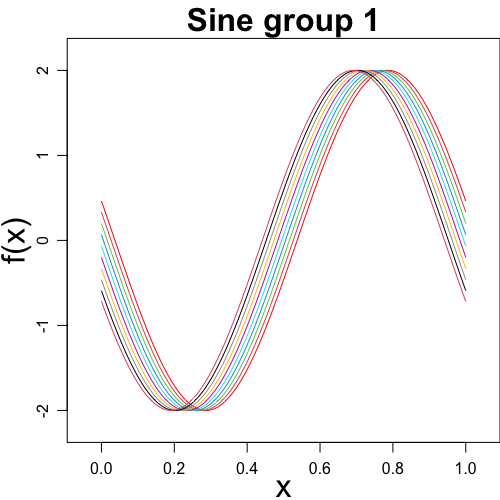}
\includegraphics[width=2.5in]{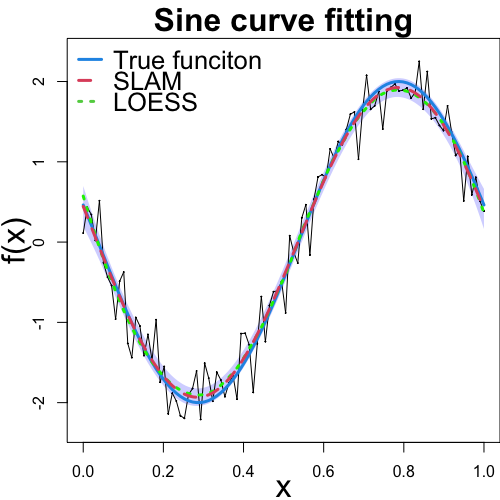}
\includegraphics[width=2.5in]{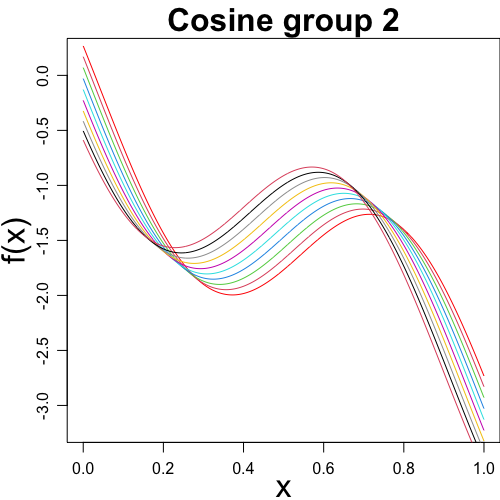}
\includegraphics[width=2.5in]{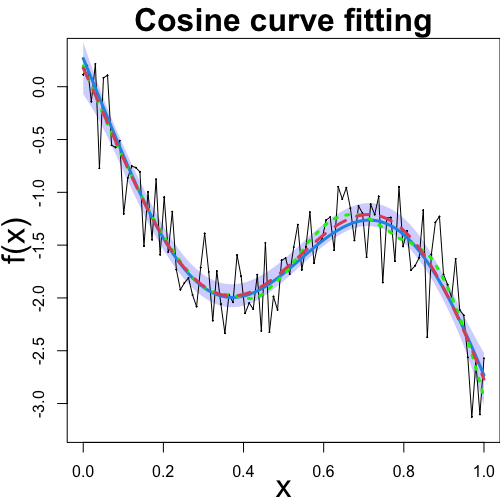}
\caption{{\bf Simulation study.} (Top) The ten simulated regression functions and corresponding curve fitting, for the first subject only, for group 1 with sine functions, and (Bottom) for group 2 with cosine functions. In the curve fitting plots, the light blue shaded area indicates the 95\% credible interval for $f(x)$, estimated from the sample paths $\{\hat{f}^{(d)}(x)\}_{d=1}^D$ drawn from the posterior derivative Gaussian process, and corresponding to the sampled latencies $\{\t^{(d)}\}_{d=1}^D$. The mean of those realizations is shown as the red dashed line, the green dashed fitted curve is obtained from the LOESS method, and the true $f(x)$ is shown in blue color.}
\label{fig:reg_fcn}
\end{figure}

For subject $s = 1, 2, \dots, 10$, group $g = 1, 2$, the simulated regression functions of the two groups are 
\begin{align*}
f(x)_{1s} = -2\sin(2 \pi x + s/15 - 0.3),
\end{align*}
and 
\begin{align*}
f(x)_{2s} = \cos(2\pi x + s/10 + 1.2) - 3x,
\end{align*}
respectively, where $x \in \mathcal{X} = (0, 1)$. The regression functions are shown in Figure \ref{fig:reg_fcn}. The simulated data are generated from $y_{igs} = f(x_i)_{gs} + \epsilon_{igs}$, $i = 1, 2, \dots, n$, $g = 1, 2$, where $x_i = (i-1)/99$ and $\epsilon_{igs} \iid N(0, 0.25^2)$ and $n=100$.

Both groups have two components in their waveforms. Note that in group 1 where data are generated from the sine functions, all the curves have the same amplitude size and only latency changes, accounting for individual differences. If zero is the baseline, the amplitude is two and negative two for the peak and dip for all the subjects. The latency for the dip or the first ERP component ranges from 0.19 to 0.29, and the latency for the peak or the second ERP component ranges from 0.69 to 0.79. The curves in the second group, on the other hand, have their own latency and amplitude. For the dip, its latency ranges from 0.23 to 0.37, and its amplitude changes from -1.57 to -2. For the peak, its latency ranges from 0.57 to 0.71, and its amplitude changes from -0.83 to -1.26. For each curve, 100 observed data points are generated by adding mean zero Gaussian errors with a standard deviation 0.25. The simulation is set to mimic real ERP data. 

\subsection{Parameter Settings}
We fit our novel SLAM with one-factor ANOVA to the entire data containing 20 ERP trajectories by setting $M=2$ components in both groups, and assuming that one stationary point is in $(0, 0.5)$, and the other in $(0.5, 1)$. The initial values of the MCEM algorithm are specified as follows. The subject or individual level latency is randomly drawn from the uniform distribution, all the beta coefficients are set to zero and the parameters $\bet$ and $\sigma$ are set to one. In each E-step, the first 100 draws are considered burnin, and then further 2000 MCMC iterations, with no thinning, are saved for estimation, with 500 subsampled draws used for the subsequent M-step, to optimize the hyperparameters of the GP kernel. The iterative algorithm stops when the current updated parameters and the ones of the previous iteration have a difference less than $10^{-5}$. The final 20,000 MCMC samples are saved, and the posterior distribution of the stationary points is summarized based on the MCMC samples. To do the inference for group-level latency, the logit link function is used for transformation. When comparing with the LOESS-based method we the \texttt{loess.wrapper()} function in the R package \texttt{bisoreg} and the function \texttt{PeakDetection()} provided in the online supporting information of \cite{hasenstab2015identifying} at \url{https://onlinelibrary.wiley.com/doi/10.1111/biom.12347} to each individual curve to estimate the underlying ERP waveform, latency and amplitude. The LOESS smoothing parameter was determined by 5-fold cross-validation from a sequence of values from 0.25 to 1 with increments 0.05. The dip searching window is set to be $(0.01, 0.5)$, and the peak time window is $(0.5, 0.99)$. The method provides point estimates with no uncertainty quantification. 

\subsection{Results on Latency Estimation}
\begin{figure}
\centering
\includegraphics[width=2.5in]{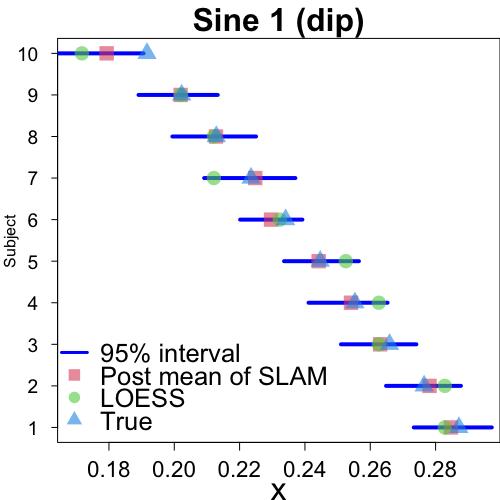}
\includegraphics[width=2.5in]{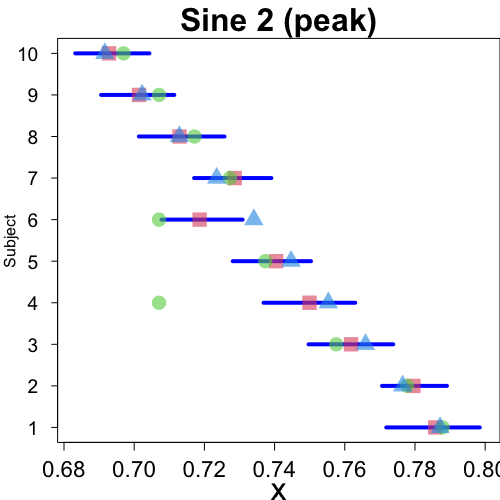}
\includegraphics[width=2.5in]{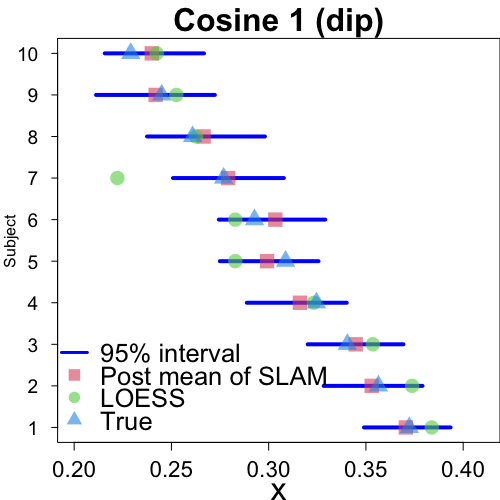}
\includegraphics[width=2.5in]{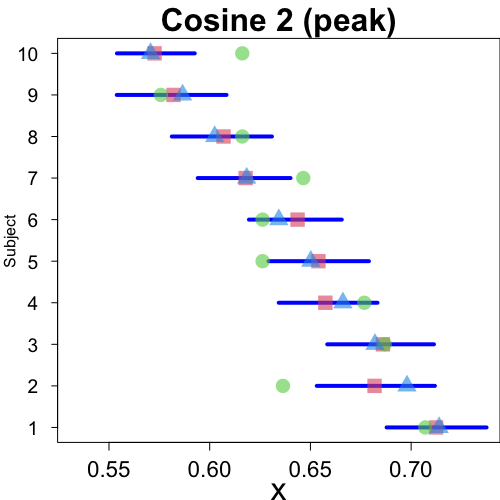}
\caption{{\bf Simulation study:} (Top) Latency estimates for group 1 with sine functions and (Bottom) for group 2 with cosine functions.} \label{fig:lat_logit}
\end{figure}

The latency estimates from SLAM, for all subjects, calculated as posterior means, are shown in Figure \ref{fig:lat_logit}, together with the 95\% credible intervals. For comparison, the LOESS-based estimates are also shown. Results show that, even though some of the LOESS-based estimates are closer to their corresponding true stationary points than the posterior mean estimates derived from our algorithm, there is more variability in the LOESS-based estimates, with more of the estimates away from the true values. This unstable phenomenon is more significant for group 2, with the cosine functions. This is confirmed by the root mean square error (RMSE), calculated as $\sqrt{\frac{\sum_{s=1}^{10}\sum_{g=1}^2\sum_{m=1}^2(\hat{t}_{gs}^m - t_{gs}^m)^2}{40}}$, as shown in Table \ref{tbl:rmse}, which reports RMSEs averaged over 100 replicates, with standard deviations in parentheses. Indeed, the proposed method outperforms the LOESS-based method in terms of estimation for individual data sets, as well as consistency from sample to sample. We note that the LOESS-based fit relies completelyon the sample data and requires a large sample to produce reliable estimates of the latency. In our experiment, when the simulation sample size is doubled to $n=200$ the RMSE measures, averaged over 100 replicates, are 0.009 (0.0028) for the sine group and 0.021 (0.0027), with the value in parentheses indicating standard deviation. The metrics are improved compared to the values with $n=100$. However, LOESS with $n=200$ is still outperformed by SLAM with $n=100$.

\begin{figure}[h!]
	\centering
\includegraphics[width=2.5in]{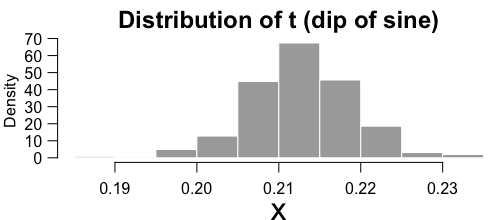}
\includegraphics[width=2.5in]{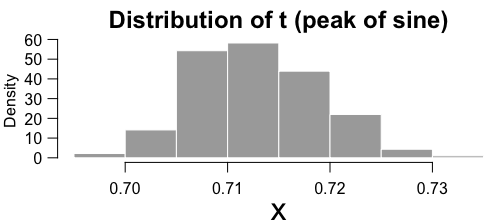}
\includegraphics[width=2.5in]{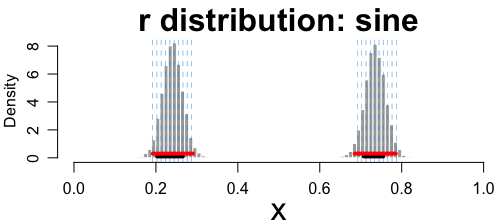}
\includegraphics[width=2.5in]{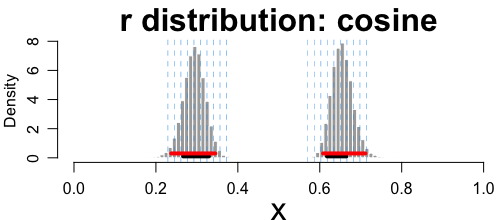}
\includegraphics[width=2.5in]{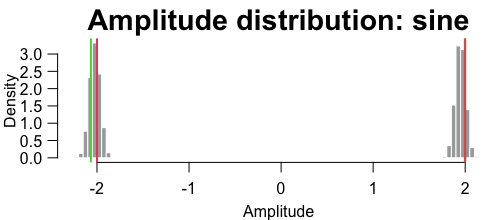}
\includegraphics[width=2.5in]{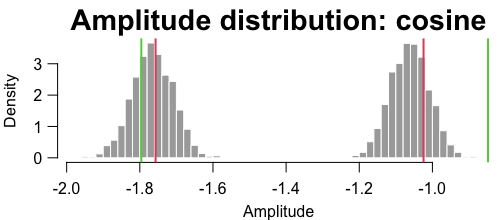}
\caption{{\bf Simulation study:} (Top) Posterior histograms of $t$ of the sine regression function for the 8th subject.
(Middle) Posterior distribution of group-level parameters $r^1$ and $r^2$ for the sine and cosine regression function settings. The blue vertical dashed lines indicate true subject-level latencies. The red segments are the 95\% credible intervals from the posterior distributions by SLAM, and the black segments are the 95\% confidence intervals using the two-step approach by fitting LOESS followed by one-way ANOVA. (Bottom) Amplitude distributions using the Max Peak method in Algorithm \ref{algo:max}. Vertical red lines indicate the true amplitudes, and the green lines indicate the LOESS estimates.	\label{fig:hist_sin}}
\end{figure}

Using the posterior median for point estimation does not change the values of the RMSEs much, due to the fact that the posterior distribution of stationary point is basically symmetric and unimodal with small variation, as shown, for one subject, in the top panel of Figure \ref{fig:hist_sin}.

\begin{table}[h!]
	\centering
\begin{tabular}{lcc}
\hline
&{\bf Latency}&\\
\hline
Method & Group 1: Sine & Group 2: Cosine \\
\hline
SLAM-posterior mean & 0.0037 (0.0000) & 0.0068 (0.0002) \\
SLAM-median & 0.0036 (0.0001) & 0.0068 (0.0002) \\
LOESS & 0.0119 (0.0021) & 0.0251 (0.0036)\\
\hline
&{\bf Amplitude}&\\
\hline
Method & Group 1: Sine & Group 2: Cosine \\
\hline
SLAM-posterior mean & 0.0421 (0.0004) & 0.0582 (0.0016) \\
SLAM-median & 0.0421 (0.0005) & 0.0581 (0.0016) \\
LOESS & 0.0661 (0.0007) & 0.0654 (0.0066)\\
\hline
\end{tabular}
	\caption{{\bf Simulation study:} Latency and amplitude comparison.  
 Average and the standard deviations (in parentheses) of the RMSEs from the 100 replicated data sets.
 }
	\label{tbl:rmse}
\end{table}

The middle panel of Figure \ref{fig:hist_sin} shows the posterior distribution of the group-level parameters $r^1$ and $r^2$ for the sine and cosine regression function settings. The 95\% credible interval for $r^1$ and $r^2$ are specified by the red segments. The original LOESS-based methods do not account for group effects and, consequently, cannot provide inference on group-level parameters. To address this limitation, we employ a two-step adaptation as in \cite{hasenstab2015identifying}, involving LOESS fitting followed by one-way ANOVA. The resulting 95\% confidence intervals are represented by black segments in Figure \ref{fig:hist_sin}. The confidence intervals appear to be narrower than the credible intervals. We also note that, while the sampling distribution of $r^1$ and $r^2$ is assumed to be Gaussian, the posterior distributions from SLAM are not necessarily Gaussian. 

As for the comparison with the DGP-MCEM method of \cite{Yu2022}, we found that the latency posterior means were almost identical to the SLAM estimates (results not shown). Also, on average, the SLAM and DGP-MCEM methods produced 95\% uncertainty bands of similar length. However, the SLAM intervals exhibited less variation than the DGP-MCEM intervals, with standard deviations 0.002 and 0.007, for the sine group, and 0.005 and 0.006 for the cosine group, respectively. More importantly, the DGP-MCEM method does not provide group-level latency estimation and has no information about the group-level latency distribution. Indeed, in \cite{Yu2022} the DGP-MCEM model is fitted separately to each group of data. With this method, however, population-level estimates can be obtained, by fitting a Gaussian mixture and calculating the normal interval $\bar{t} \pm 2 \text{sd}(\bar{t})$, where $\bar{t}$ is the mean of the Gaussian mixture and $\text{sd}(\bar{t})$ is the standard deviation of those means. We found that SLAM has a more precise estimation with a shorter interval than DGP-MCEM. Specifically, for the sine group, SLAM produces interval $(0.19, 0.29)$ for $r^1$ and $(0.69, 0.79)$ for $r^2$, compared to the normal intervals $(0.17, 0.31)$ and $(0.67, 0.80)$. For the cosine group, SLAM produces $(0.24, 0.34)$ for $r^1$ and $(0.61, 0.71)$ for $r^2$. The corresponding normal intervals are $(0.21, 0.40)$ and $(0.54, 0.74)$ for $r^1$ and $r^2$, respectively.

\subsection{Results on Amplitude}
As for amplitude, we can derive estimates from the SLAM output as follows. First, for each sampled latency, given the data, a posterior path is drawn from the posterior derivative Gaussian process.  Then, the fitted regression curve is obtained as the mean of those realizations.  Examples, for one of the subjects, are shown in Figure \ref{fig:reg_fcn}. With traditional approaches for ERP data analysis, different heuristic methods are used to estimate amplitudes, 
for example as the amplitude value of the peak of the component or as the mean/area amplitude over a selected time window or search window, such as in the LOESS-based methods \citep{Luck2005}. Since our method is able to quantify possible locations of the ERP components, or latencies, we can essentially use this information to narrow down the search window by measuring amplitudes over the range of the posterior distribution of latency. Alternatively, an even shorter search interval could be used, by considering for example any $(1-\alpha)100\%$ credible interval of the distribution of latency. Since the search window is usually set wider so as to include all individual peaks of the same component, our latency intervals provide a more precise location of those peaks. Here, to compare with the LOESS-based method, we define a peak as the largest local extremum within the search window, that is, in our case, the range of the posterior samples of latency. The method is summarized in Algorithm \ref{algo:max}. This method cares only about the largest extremum, and the shape of components does not matter.

\begin{algorithm}
	\SetAlgoLined
	For stationary point samples $\{t^{(d)}\}_{d=1}^D$, and the curve realizations $\{\hat{f}^{(d)}(x)\}_{d=1}^D$,
	\vskip 2mm
	1. Decide the range of latency $(a, b)$ of , i.e., $a = \min (t^{(1)}, \dots, t^{(D)})$ and $b = \max (t^{(1)}, \dots, t^{(D)})$\\
	2. For $d = 1, 2, \dots, D$:\\
	\begin{itemize}
		\item [(1)] Compute $\hat{f}^{(d)}(x)$ for all $x \in (a, b)$
		\item [(2)] Define $z^{(d)} := \max \hat{f}^{(d)}(x)$ for a peak or $z^{(d)} := \min \hat{f}^{(d)}(x)$ for a dip.
	\end{itemize}
	3. Collection $\{z^{(d)}\}_{d = 1}^D$ forms posterior samples of amplitude.
	\caption{Max Peak}
	\KwResult{Posterior samples of amplitude.}
	\label{algo:max}
\end{algorithm}

Finally, to examine the two methods' performance, we use the same measure, the RMSE, calculated as $\sqrt{\frac{\sum_{s=1}^{10}\sum_{g=1}^2\sum_{m=1}^2(\hat{A}_{gs}^m - A_{gs}^m)^2}{40}}$, where $A_{gs}^m$ is the true amplitude value and $\hat{A}_{gs}^m$ is its estimate, obtained as the posterior mean of our method. 
Table \ref{tbl:rmse} compares SLAM and LOESS-based methods on the amplitude estimation. The SLAM method on average has smaller RMSEs as well as smaller standard deviations. One distinct advantage of our method is that it provides uncertainty quantification. The bottom panel of Figure \ref{fig:hist_sin} shows the amplitude distribution of the sine and cosine function, for one subject, in one simulated data using the Max Peak method in Algorithm \ref{algo:max}. The distribution is Gaussian-like.

Additional investigations of this simulated setting with non-Gaussian noises and under mis-specification of $M$ can be found in the supplementary material. We find that the proposed method works well for noise terms that are not extremely far away from Gaussian. Also, while, as expected, the estimation of latency, both at group- and subject-level, is distorted, the curve fitting is somehow robust to mis-specifying $M$.
 
\subsection{Simulation from the Model} \label{sec:simdatamodel}
In this section we generate simulated data from the proposed model, and examine the ability of SLAM to capture the true parameters.

\begin{figure}[h!]
	\centering
	\includegraphics[width=2.5in]{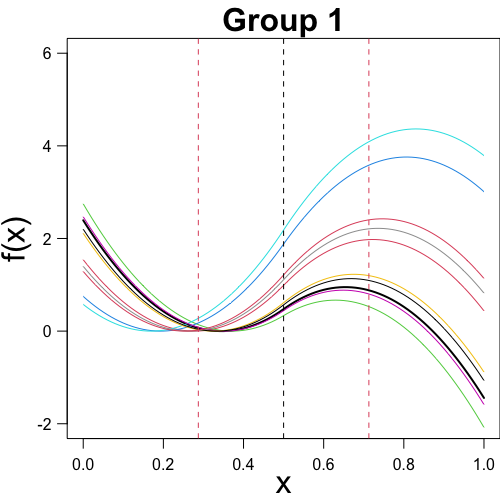}
  	\includegraphics[width=2.5in]{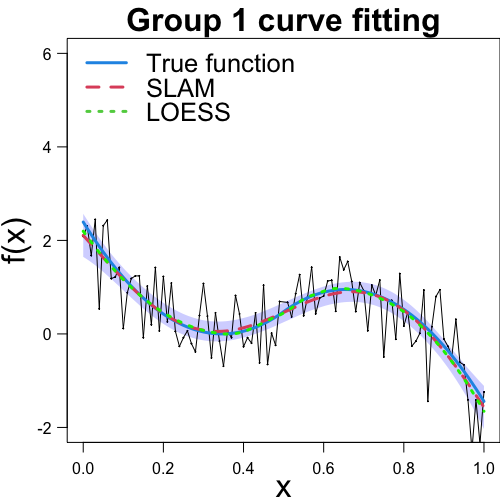}
        \includegraphics[width=2.5in]{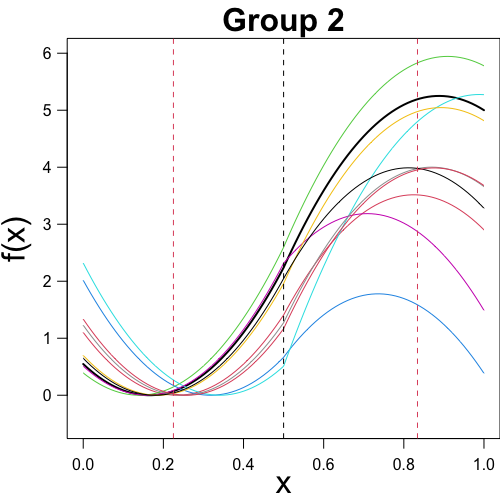}
 	\includegraphics[width=2.5in]{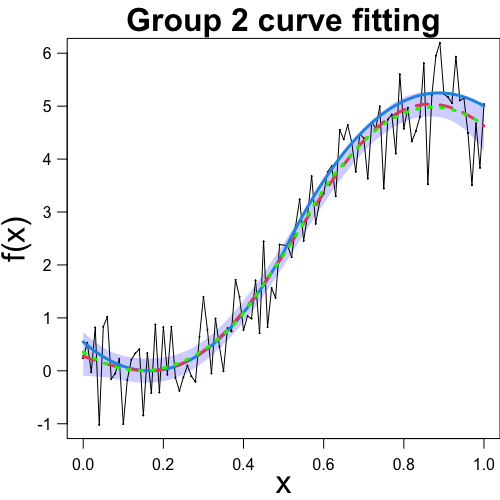}
	\caption{{\bf Simulation study:} (Top) Examples of simulated regression functions and corresponding curve fitting, for the first subject only, for group 1 with sine functions, and (Bottom) for group 2 with cosine functions. There is a group-level stationary point in the region $(0, 0.5)$ and in $(0.5, 1)$. The black dashed line separates the two regions, and the red vertical lines indicate the true group-level latencies. The true $f(x)$ is shown in blue color.}
	\label{fig:reg_fcn_model}
\end{figure}

The data are simulated from the generating process described by the proposed model. Suppose there are two ERP components and a factor with two levels being considered. The $\beta$s are set to be $\beta_0^1 = 0.3$, $\beta_0^2 = -0.3$, $\beta_1^1 = -0.5$, and $\beta_1^2 = 1$. Through the logit link, the true group level latencies are $r_1^1 = 0.57$, $r_1^2 = 0.43$, $r_2^1 = 0.45$, $r_2^2 = 0.67$. Each group has 10 subjects and for each subject $s$ the subject-level latencies are generated from 
$$(t_{gs}^m \mid r_{g}^m, \eta_{g}^m = 8) \iid \text{gbeta}\left(8r_{g}^m, 8(1 - r_{g}^m), \, a^{m}, \, b^{m}\right),$$
where $(a^1, b^1) = (0, 0.5)$ and $(a^2, b^2) = (0.5, 1)$. The regression function for $x \in (0, 0.5)$ is set to be $f_{gs}^1(x) = 20 (x - t_{gs}^1)^2$, and the one for $x \in (0.5, 1)$ is $f_{gs}^2(x) = 20 \left[ -(x - t_{gs}^2)^2 + (b^1 - t_{gs}^1)^2 + (a^2 - t_{gs}^2)^2\right]$. The true regression function defined in $(0, 1)$ is the one with $f_{gs}^1(x)$ and $f_{gs}^2(x)$ combined. Data of size $n=100$ are generated by adding Gaussian random noise $N(0, \sigma^2)$, with $\sigma=0.5^2$ set to a level similar to the noise variation in the real ERP data used in the next section. The regression functions are shown in Figure \ref{fig:reg_fcn_model}.

For prior specification, we assign standard normal priors to all $\beta$s and impose an $Ga(1/2, 1/2)$ on all $\eta$s and an $IG(1/2, 1/2)$ on $\sigma^2$. To implement the algorithm, the initial values of $t_{gs}^1$ are randomly drawn from the uniform distribution $\mathcal{U}(0, 0.5)$ and those for $t_{gs}^2$ from $\mathcal{U}(0.5, 1)$. The initial values of all $\beta$s are set to 0, and those for the $\eta$s and $\sigma^2$ to 1. The variance of the proposal distribution is updated every 40 iterations, so that the acceptance rate of the M-H steps is around 35\%. The posterior samples are such that Gelman and Rubin’s potential scale reduction factors or $\widehat{R}$ is less than 1.1 for all parameters. It takes about 800 seconds to obtain a posterior sample of size 10,000.

Examples of curve fitting, for one subject, are shown in Figure \ref{fig:reg_fcn_model}, while Table \ref{tbl:par_est} summarizes the results for the group-level latencies, $\beta$s and $\sigma$. Mean and standard deviations of posterior mean and RMSE are computed based on 50 replicated data sets. The parameters are well estimated by SLAM, and the inference improves when more subjects are under study. Plots of the posterior distributions of $\beta$s and $r$s, as well as median and MAP estimates are reported in the supplementary material.

\begin{table}[h!]
	\centering
\begin{tabular}{cccccc}
\hline
& &\multicolumn{2}{c}{Mean (SD) of posterior mean} & \multicolumn{2}{c}{Mean (SD) of RMSE}\\
\cline{3-4}
\cline{5-6}
Parameter & True value & 10 subjects & 20 subjects & 10 subjects & 20 subjects \\
\hline
$r_1^1$ &  0.57    & 0.54 (0.033)     & 0.56 (0.027)    & 0.077 (0.016)  & 0.050 (0.012)      \\
$r_1^2$ &  0.43    & 0.46 (0.028)     & 0.44 (0.024)    & 0.077 (0.012)  & 0.051 (0.011)      \\
$r_2^1$ &  0.45    & 0.45 (0.035)     & 0.45 (0.029)    & 0.075 (0.010)  & 0.051 (0.009)      \\
$r_2^2$ &  0.67    & 0.63 (0.033)     & 0.65 (0.025)    & 0.084 (0.017)  & 0.052 (0.010)      \\
$\beta_0^1$ & 0.3  & 0.17 (0.135)     & 0.24 (0.109)    & 0.314 (0.064)  & 0.205 (0.049)      \\
$\beta_0^2$ & -0.3 & -0.17 (0.114)    & -0.24 (0.100)    & 0.317 (0.049)  & 0.210 (0.043)      \\
$\beta_1^1$ & -0.5 & -0.38 (0.276)    & -0.44 (0.222)    & 0.463 (0.115)  & 0.319 (0.100)      \\
$\beta_1^2$ &  1   & 0.72 (0.244)     & 0.88 (0.203)    & 0.538 (0.127)  & 0.351 (0.093)      \\
$\sigma$    & 0.5  & 0.57 (0.003)     & 0.57 (0.003)    & 0.074 (0.002)  & 0.074 (0.002)      \\
\hline
\end{tabular}
	\caption{{\bf Simulation study:} Parameter estimation. The mean and standard deviation of posterior mean and RMSE are computed from 50 replicates of data with size 100.}
	\label{tbl:par_est}
\end{table}

\section{ERP Data Analysis}\label{sec:erp}
We now perform an analysis of real ERP data on speech recognition, where we assess the effect of age on two ERP components of interest. We provide subject- and group-level estimates of latencies and amplitudes of these characteristic components, with uncertainty quantification, and discuss comparisons with the inference provided by the method of \cite{Yu2022}. 

\subsection{Data}
\begin{figure}[h!]
	\centering
\includegraphics[width=3.8in, height=2.6in]{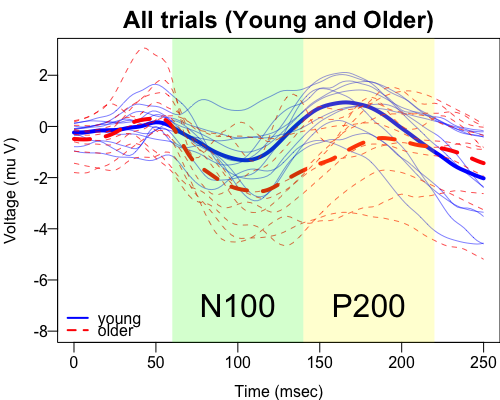}
\includegraphics[width=2.5in]{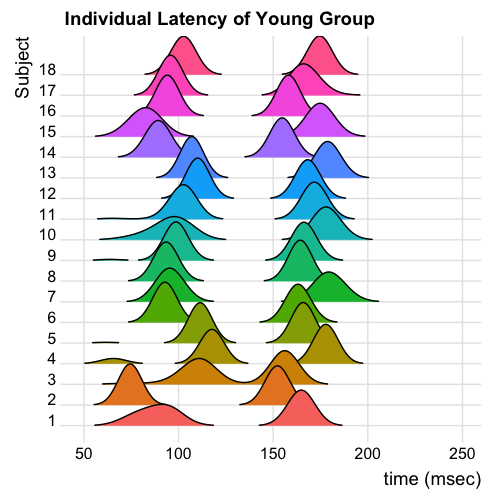}
\includegraphics[width=2.5in]{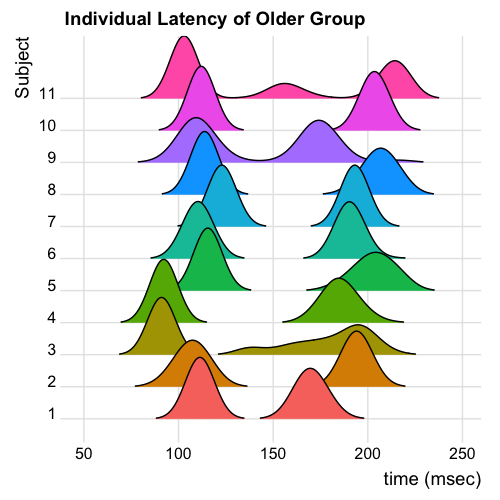} 
 \caption{{\bf Case study:} (Top) Subject-level ERP curves and grand average ERP curves. Solid: Young group. Dashed: Older group. The 0 ms time, which corresponds to time point 100, is the start of the onset of sound. The N100 component of interest is the dip characterizing the signal in the time window [60-140] msec colored in green. The P200 component characterizes the time window [140-220] msec, colored in yellow. (Bottom) Posterior distribution of subject-level latencies.}
	\label{fig:erp_data}
\end{figure}

ERP data were collected from an experiment on speech recognition conducted at Rice University \citep{Noe2019} and analyzed in \cite{Yu2022}. The experiment involved 18 college-age students and 11 older controls, as aging is known to cause difficulties in perceiving speech, especially in noisy environments \citep{Peelle2016}. The older subjects have ages ranging from 47 to 91 years old with a mean age of 67.78 years \citep{Noe2022}. EEG signals were recorded continuously during the speech task and standard preprocessing \citep{Luck2005} was done using the ERPLAB toolbox \citep{LopezCalderon2014} in EEGlab \citep{Delorme2004} before the data analysis. The experiment resulted in a total of 2304 trials. The top panel of Figure \ref{fig:erp_data} displays the ERP waveforms for all subjects, averaged across all trials and six electrodes. 

Common methods of analyzing ERP data involve averaging ERP waveforms over a particular condition and window of interest, to obtain the magnitudes and latencies of specific components. Specific ERP components are expected to be associated with speech perception (Tremblay et al., 2003). In particular, the N100 component, which captures phonological (syllable) representation, is of interest. This component is identified by the latency of the negative deflection in the time window [60, 140] ms after the stimulus onset. The specific starting point of the N100 window is typically chosen by visually inspecting the grand average data, to obtain a window where the ERP curve has a negative value. In addition to the N100, we also considered the P200 component, in the window [140, 220] msec, an auditory feature representing higher-order perceptual processing, modulated by attention. Recent studies have identified the P200 time window as potentially critical for processing higher order speech perception. 

For posterior inference, we generated 21000 MCMC draws with 1000 burn-ins and thinned the chain by keeping every 10th draw, obtaining 2000 posterior samples.  The final MCMC simulation in the MCEM algorithm was monitored and stopped when the potential scale reduction factor of every parameter was below 1.1, reaching approximate convergence. The full MCMC diagnostics are reported in the supplementary materials.

\subsection{Inference on Latency}
The bottom panel of Figure \ref{fig:erp_data} illustrates the posterior density of the subject-level latency parameters in both young and older groups, based on 2000 posterior samples produced by the MC procedure in the E-step. We observe higher variability in the latency distribution for the older group, especially in the period after N100. Plots show substantial subject-level variability, as also noticed by \cite{Yu2022}. Overall, most of the subjects show latencies nicely concentrated around the two components of interest, with a few showing larger uncertainty, and one (subject 11 in the older group) having a third smaller latency, suggesting a possible outlier.

\begin{figure}[h!]
	\centering
        \includegraphics[width=1.85in]{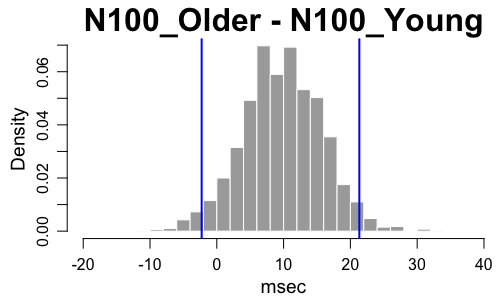}
        \includegraphics[width=1.85in]{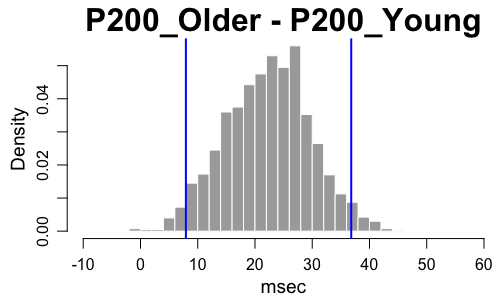}
        \includegraphics[width=1.85in]{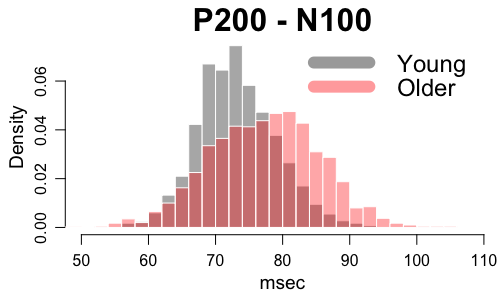}
	\includegraphics[width=2.8in]{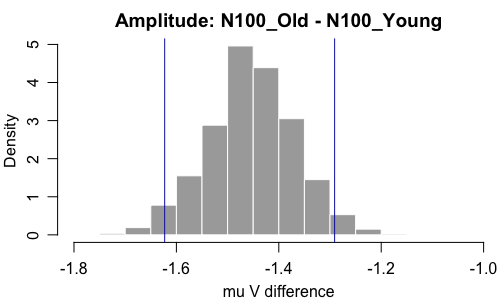}
	\includegraphics[width=2.8in]{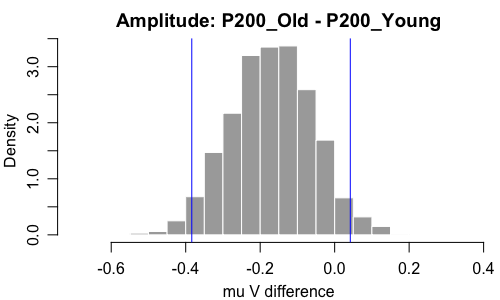}
	\caption{{\bf Case study:} (Top) Latency difference between groups and latency shift effect on the right. (Bottom) Amplitude difference between groups.}
	\label{fig:case_inf}
\end{figure}

Furthermore, we can obtain a probabilistic statement on whether the ERP peaks of the older group occur later than the peaks of the young group by calculating the posterior distribution of the difference in latency between groups. This is shown in Figure \ref{fig:case_inf}. In general, the peaks of N100 and P200 occur on average with a delay of 11.02 and 21.72 msec, respectively, in the older group. Also, the probability that the latency difference of N100 is greater than zero is $95.9\%$ and that of P200 is $98.5\%$, suggesting significant latency differences between the two groups. These insights align with established results in aging theory, as these components rely on inhibitory connections to resolve perceptual objects and older adults are thought to have generally less inhibition and slower resolution of objects \citep{Yi2014, Tremblay2003}.

Further probabilistic evidence on the latency shifting effect is provided by the distribution of $r^2 - r^1$, see Figure \ref{fig:case_inf}. For example, for the young group, the probability that $r_{young}^2 – r_{young}^1$ is between 70 and 80 msec is about $56.4\%$, and that of the older group is about $41.9\%$. Notice that the time lag between N100 and P200 in the young group is shorter than the lag in the older group which has a higher chance of a time lag longer than 90 msec.

\subsection{Inference on Magnitude}

As with latency estimation, magnitude estimation can be performed at both group level and subject level. In particular, the posterior samples of the ERP curves can be used to quantify the uncertainty about amplitudes of N100 and P200 by integrating the estimated curve over the range of latency, say $(a, b)$, and finding the time point or latency $t$ such that the integral is at half of its full value. The $50\%$ area latency method or the half-integral method shown in Algorithm \ref{algo:integral} is more robust to noise, compared with the basic point estimate peak amplitude \citep{Luck2012}. In addition to Max Peak and Half-integral Peak, other methods such as measuring mean voltage over a given time window can be used to compute the amplitude using the posterior samples within our modeling framework.  We note that it is common to use zero micro voltage as the reference, or baseline, to compute the amplitude. However, the positive peak at P200 of the older group is all below zero micro voltage. Therefore, to make the amplitude comparable, we use the peak of N100 as the baseline for the calculation of the amplitude of P200. Therefore, the amplitude shows the increase in units of micro voltage after N100, in response to the stimulus of the experiment.
	
\begin{algorithm}
	\SetAlgoLined
	\vskip 2mm
	1. Decide the range of latency $(a, b)$ of samples $\{r^{(d)}\}_{d=1}^D$, i.e., $a = \min (r^{(1)}, \dots, r^{(D)})$ and $b = \max (r^{(1)}, \dots, r^{(D)})$\\
	2. For $d = 1, 2, \dots, D$:\\
	\begin{itemize}
		\item [(1)] Find $t^{(d)} \in (a, b)$ such that $\int_{a}^{t^{(d)}} \hat{f}^{(d)} \,dt = \frac{1}{2}\int_{a}^{b} \hat{f}^{(d)} \,dt$.
		\item [(2)] Compute and define $z^{(d)} := \hat{f}^{(d)}(t^{(d)})$. 
	\end{itemize}
	2. Collection $\{z^{(d)}\}_{d = 1}^D$ forms posterior samples of amplitude.
	\caption{Half-Integral Peak}
	\KwResult{Posterior samples of amplitude of $N100$ and $P200$.}
	\label{algo:integral}
\end{algorithm}

Figure \ref{fig:case_inf} shows that the amplitude size of N100 for the older group is significantly larger than the amplitude for the young group. We preserve the negative sign to indicate that the peak is below zero. For P200, the peak measurements are greater than zero since the baseline is at the peak of N100. The 95\% credible interval for \texttt{P200\_Old} - \texttt{P200\_Young} includes zero, showing a weak significance of the P200 amplitude difference. Nevertheless, with about 80\% probability, the P200 amplitude for the young group is larger than that of the older group.

We conclude by noting that, with respect to the method proposed in \cite{Yu2022}, the hierarchical structure of SLAM allows learning the latency parameters at the subject and group level in a single model. \cite{Yu2022} estimate subject-level latencies only, and the group-level latencies are estimated separately by fitting a Gaussian mixture on the posterior samples of $t$, of all participants. With their approach, the 95\% confidence intervals of the N100 group-level latencies are $[85.38, 114.76]$ and $[90.38, 126.67]$, for the young and older group, respectively. With SLAM, the 95\% credible intervals for $r^1$ are $[86.73, 104.45]$ and $[98.08, 117.82]$ for the young and older group, respectively. The SLAM intervals are generally shorter and provide more precise interval estimation.

As for the other parameters, posterior means and 95\% credible intervals are 0.49 (0.48, 0.51) for $\sigma$, -0.13 (-0.57, 0.23) for $\beta_0^{young}$, -0.59 (-0.92, -0.26) for $\beta_0^{older}$, 0.50 (-0.11, 1.10) for $\beta_1^{young}$, and 1.17 (0.41, 2.04) for $\beta_1^{older}$.

\section{Discussion} \label{sec:conclude}
In this paper, we have proposed SLAM, a novel Bayesian approach for the estimation of the amplitude and latency of ERP components.
The novel SLAM is a unified framework and integrative approach that enhances the DGP model proposed in \cite{Yu2022} and offers comprehensive statistical inference about the parameters of interest in ERP studies. As for the method of \cite{Yu2022}, SLAM estimates the uncertainty about latency and hence provides a data-driven model-based time window for measuring the magnitude of ERP components. While traditional methods provide a point estimate of magnitude, we further exploit our approach to obtain posterior samples of magnitude. Moreover, SLAM uses the fitted smooth ERP curve and therefore washes out noise that tends to affect the point estimate of a peak, especially when the Max Peak method is used.

In addition, SLAM incorporates a latent ANOVA structure that allows to examine how factors or covariates affect the magnitude and/or latency of ERP components, which is the main interest in ERP research as magnitude and latency reflect cognitive, psychological or neural processes.  
While traditional methods require a two-step approach or several methods to complete the statistical analysis, SLAM wraps up everything in one single model, producing inference by posterior samples of the group latency/magnitude difference. It examines not only the subjects' individual differences but also group-level differences that facilitate comparing different characteristics or factors, such as age in our illustration. Our results have shown that ERP components N100 and P200 are delayed in older adults, and that the N100 amplitude is generally larger for adults than for the younger group (with a negative sign).

Our model lays the foundation for more sophisticated statistical modeling for ERP analysis. As with generalized linear models and neural networks, SLAM can accommodate various valid link functions or activation functions with input domain $(0, 1)$. Besides the logit function, which we have used in our analyses, other popular link functions for a variable between zero and one include the probit and complementary log-log (cloglog) functions. Both logit and probit are symmetric functions, and cloglog is asymmetrical. In our explorations on simulated data, we find that SLAM is robust to the choice of link functions, as different link functions produced nearly the same posterior distribution of the parameters, leading to similar RMSE of latency and amplitude, as well as to similar credible intervals. This robustness adds to the flexibility of our model. When the interest is on how latency changes with factor levels, that is the $\beta$ coefficients, we recommend using the logit function for ease of interpretation, as the coefficients are related to the change in log odds. Furthermore, although ERP studies focus on how categorical factors affect amplitude and latency, numerical covariates such as blood pressure and weight can be included in our latent regression hierarchy. Other extensions of the proposed method could include treatments of other noise distributions accounting, for example, for autoregressive correlation or trial variability. Spatial modeling and mapping on multiple electrodes could be other possible extensions.

\section{Data Availability Statement}
Behavioral and Event-Related Potentials data that support the findings in this paper are available on PsyArxiv \citep{Noe2019} at https://osf.io/c7k4s/ (DOI 10.17605/OSF.IO/C7K4S).

\section{Statements and Declarations}
The authors have no relevant financial or non-financial interests to disclose. The authors have no competing interests to declare that are relevant to the content of this article. All authors certify that they have no affiliations with or involvement in any organization or entity with any financial interest or non-financial interest in the subject matter or materials discussed in this manuscript. The authors have no financial or proprietary interests in any material discussed in this article.

\section*{Appendix: Monte Carlo sampling in the E-step of the MCEM algorithm}
\label{sec:app}

In the Monte Carlo E-step of the proposed MCEM Algorithm \ref{algo:MCEM} for SLAM, we simulate $\t_{gs}$, $\beta_0^{m}$, $\beta_a^{m}$, $\eta_g^{m}$ and $\sigma^2$ using the following conditional distributions and Metropolis-Hasting steps.

\begin{itemize}
\item With the property that $\y_{gs}$ are conditionally independent given $\t_{gs}$ and $\hat{\btheta}^{(j)} $, the conditional density of $\t_{gs}$ is 
\begin{align*}
p\left(\t_{gs} \mid \y_{gs}, \r_{g}(\bbeta), \sigma^2, \hat{\btheta}^{(j)} \right) = p\left(\y_{gs} \mid \t_{gs}, \sigma^2, \hat{\btheta}^{(j)} \right)\prod_{m=1}^M \pi(t_{gs}^m \mid r_{g}^m(\bbeta_0, \bbeta_1^m), \eta_{g}^m)\\
= N\left(\bmu_{gs}, \bSig_{gs}+\sigma^2\I \right) \prod_{m=1}^M \text{gbeta}(t_{gs}^m \mid r_{g}^m(\bbeta_0, \bbeta_1^m), \eta_{g}^m, a^{m}, b^{m})
\end{align*}
and a Metropolis-Hastings (M-H) sampling is performed. The simple independent symmetric uniform proposal $(t_{gs}^m)^* \ind \text{Unif}(a^{m}, b^{m})$ could be used. For more efficient sampling, we can also use the truncated normal distribution as the proposal distribution $(t_{gs}^m)^* \sim TN\left( (t_{gs}^m)_{(t-1)}, (C_{gs}^m)^2, a^{m}, b^{m}\right)$ with support $(a^{m}, b^{m})$, where $(t_{gs}^m)_{(t-1)}$ is the $(t-1)$th draw of $t_{gs}^m$ in the E-step, and $(C_{gs}^m)^2$ is the tuning variance that keeps the user-defined acceptance rate, 35\% for example.

\item  The conditional distribution of $\beta_0^{m}$ is given by
\begin{align*}
p\left(\beta_0^{m} \mid \t, \bet \right) \propto \left[ \prod_{g=1}^{G}\prod_{s=1}^{S_{g}} \text{gbeta}(t_{gs}^m \mid \{ \beta_{a}^m \}_{a=1}^{G-1}, \beta_0^m,\eta_{g}^m, a^{m}, b^{m})\right] \\
N\left(\beta_0^m \mid \mu_0^m, (\delta^m_0)^2 \right).
\end{align*}
A Metropolis step is done with Gaussian proposal $(\beta_0^{m})^* \sim N((\beta_0^{m})_{(t-1)}, C_0^{m})$, where $(\beta_0^{m})_{(t-1)}$ is the $(t-1)$th draw of $\beta_0^{m}$ in the E-step, and $C_0^{m}$ is the tuning variance that keeps the acceptance rate at around 30\%.

\item The conditional distribution of $\beta_{a}^{m}$ is given by
\begin{eqnarray*}
p\left(\beta_{a}^{m} \mid \t, \bet \right) \propto
\left[ \prod_{g=1}^{G}\prod_{s=1}^{S_{g}} \text{gbeta}(t_{gs}^m \mid \{ \beta_{a}^m \}_{a=1}^{G-1}, \beta_0^m,\eta_{g}^m, a^{m}, b^{m})\right]\\
N\left(\beta_{a}^m \mid \mu_1^m, (\delta^m_1)^2 \right).
\end{eqnarray*}
A Metropolis step is done with the Gaussian proposal $(\beta_{a}^{m})^* \sim N((\beta_{a}^{m})_{(t-1)}, C_1^{m})$, where $(\beta_{a}^{m})_{(t-1)}$ is the $(t-1)$th draw of $\beta_{a}^{m}$ in the E-step, and $C_1^{m}$ is the tuning variance that keeps the acceptance rate at around 30\%. 

\item The conditional distribution of $\eta_{g}^m$ is 
\begin{align*}
p\left(\eta_{g}^m \mid \t_{g}^m, r_{g}^m(\bbeta_0, \bbeta_1^m) \right) = \left[ \prod_{s=1}^{S_{g}} \text{gbeta}(t_{gs}^m \mid r_{g}^m, \eta_{g}^m, a^{m}, b^{m})\right] Ga\left(\eta_{g}^m \mid \alpha_{\eta}, \beta_{\eta} \right),
\end{align*}
and a regular random walk proposal on the logarithm of $\eta_{gl}^m$ is used for sampling.

\item  A Gibbs step is done on $\sigma^2$ by drawing from its conditional density
\begin{align*}
p\left(\sigma^2 \mid \y, \t, \hat{\btheta}^{(j)} \right) = \left[\prod_{g= 1}^G \prod_{s=1}^{S_{g}} N\left(\bmu_{gs}, \bSig_{gs} +\sigma^2\I \right) \right]IG(\sigma^2 \mid \alpha_{\sigma}, \beta_{\sigma}), 
\end{align*}
which is an inverse gamma distribution with shape parameter $\dfrac{n}{2}(\sum_{g=1}^G S_{g}) + \alpha_{\sigma}$ and scale parameter $\left(\frac{1}{2}\sum_{g=1}^G\sum_{s=1}^{S_{g}}\y_{gs}'A_{gs}^{-1}\y_{gs}\right)+\beta_{\sigma}$, where $$A_{gs} = \tau_0^2 \left(k_{00}(\bm{x}, \t_{gs}) - k_{01}(\bm{x}, \t_{gs})k_{11}^{-1}(\t_{gs}, \t_{gs})k_{10}(\t_{gs}, \bm{x})\right) + \I$$ and $\tau^2 = \tau_0^2\sigma^2$.

\end{itemize}

\nocite{*}
\bibliographystyle{tfcse}
\bibliography{slam}

\end{document}